\documentclass[a4paper,ngerman,twoside,notitlepage]{article}
\usepackage{float}
\usepackage{makeidx}
\usepackage{amsmath}

\input{tcilatex}
\newcommand{\vekt}[1]{\mbox{\boldmath $#1$\unboldmath}}

\begin{document}

\title{Problem with the derivation of the Navier-Stokes equation\\
by means of Zwanzig-Mori technique: \\
Correction and solution}
\author{J. Piest}
\maketitle

\begin{abstract}
The derivation of the Navier-Stokes equation starting from the Liouville
equation using projector techniques yields a friction term which is
nonlinear in the velocity. As has been explained in the 1. version of this
paper, when the second-order part of the term is non-zero, this leads to an
incorrect formula for the equation.

In this 2. version, it is shown that the problem is due to an inadequate
treatment of the correlation function $K_{2}$. Repeating the calculation
leads to zero second-order part. The Navier-Stokes equation is correctly
derived by projection operator technique.
\end{abstract}

\section{Introduction}

The derivation of hydrodynamic equations by Zwanzig-Mori technique is a
method well-established in the literature. In version 1 of this paper, the
author reported a problem which occurred when the Navier-Stokes order of the
momentum equation is considered: As is well known, the non-dissipative part
shows to be of second order in the fluid velocity, $\mathbf{u}$. The
dissipative part is also non-linear in $\mathbf{u}$. In order to obtain the
correct form of the Navier-Stokes equation, it is necessary for the
second-order part of the dissipation term to vanish. As far as the author
knows, this aspect has not been investigated earlier. The calculation in
version 1 yielded a second-order part different from zero.

In the present version it is shown that the problem resulted from an
incomplete formula for the correlation function $K_{2}$. Actually, the
second-order term is zero. Thus, the Navier-Stokes equation is correctly
derived by projection operator technique.

\bigskip

\section{Hydrodynamic equations}

\bigskip

For the basic definitions of microscopic variables, readers are referred to %
\cite{pi07-4}. In order to keep the paper reasonably self-contained, the
results of the Zwanzig-Mori analysis are copied from there. The analysis is
for incompressible constant density/temperature processes. The hydrodynamic
momentum equation is a specialization of the general mean value equation
(3.1):

\begin{equation}
\rho (\frac{du_{a}}{dt}+u_{c}\nabla _{c}u_{a})=-\nabla _{a}P+D_{\alpha }
\label{2.1}
\end{equation}

In\ (3.15) this equation is given for stationary processes. Latin indices
run from 1 to 3. $\rho $\ is the fluid density; $\mathbf{u}$\ as well as the
pressure $P$\ and the dissipative force $\mathbf{D}$\ depend on space and
time. For the latter the formula is obtained:

\begin{equation}
D_{a}(\mathbf{x},t)=\beta \nabla _{c}\int_{0}^{t}dt^{\prime }\int d\mathbf{x}%
^{\prime }R_{abcd}(\mathbf{x},t,\mathbf{x}^{\prime },t^{\prime })\nabla
_{d}^{\prime }u_{b}(\mathbf{x}^{\prime },t^{\prime })  \label{2.2}
\end{equation}

with the kernel function:

\begin{equation}
R_{abcd}(\mathbf{x},t,\mathbf{x}^{\prime },t^{\prime })=\langle \mathcal{[G}%
(t^{\prime },t)\hat{s}_{ac}(\mathbf{x},t)]\hat{s}_{bd}(\mathbf{x}^{\prime
},t^{\prime })\rangle _{L,t^{\prime }}  \label{2.3}
\end{equation}

$\beta $ is the inverse kinetic temperature. $\langle \rangle _{L,t}$\
denotes the expectation with respect to local equilibrium, cf. \cite{pi07-4}%
\ (3.4) to (3.6). $\hat{s}_{ac}$\ is the projected momentum flux density:

\begin{equation}
\hat{s}_{ac}(\mathbf{x},t)=(1-\mathcal{P}(t))s_{ac}(\mathbf{x})  \label{2.4}
\end{equation}

$s_{ac}$\ is the momentum flux density. $\mathcal{P}(t)$\ is the
Zwanzig-Mori projection operator; for any phase space function $g$, it is
defined:

\begin{equation}
\mathcal{P}g=\langle g\rangle _{L}+\langle g\,\delta a\rangle _{L}\ast
\langle \delta a\,\delta a\rangle _{L}^{-1}\ast \delta a  \label{2.5}
\end{equation}

For shortness, the time parameters have been omitted here. $\ast $ denotes
an operation which consists of a product, a summation over 5 index values
and a space integration. $a$\ are the microscopic densities of the conserved
quantities. $\delta a=a-\langle a\rangle _{L}$. $\langle \rangle ^{-1}$
denotes the inverse matrix. $\mathcal{G}(t^{\prime },t)$\ is a time-ordered
exponential operator:

\begin{equation}
\mathcal{G}(t^{\prime },t)=\exp _{-}\{\int_{t^{\prime }}^{t}dt^{\prime
\prime }\,\mathcal{L}(1-\mathcal{P}(t^{\prime \prime }))\}  \label{2.6}
\end{equation}

$\mathcal{L}$ is the Liouville operator, cf. \cite{pi07-4}\ (2.4).

\section{2$^{nd}$ order term of the friction force}

\setcounter{equation}{0}

In this paper, we consider the second-order verlocity approximation of the
friction Term $\mathbf{D}$. This should not be confused with the
second-order wave number approximation of the linear part of $\mathbf{D}$
which leads to the well-known Stokes form of that term. This latter
approximation is not considered here.

As is explained in \cite{pi07-4}, the 2$^{nd}$ order part $\mathbf{D}^{(2)}$%
\ of the dissipative force is obtained from (\ref{2.2}) by inserting the
linear part $R^{(1)}$\ of the kernel function given by \cite{pi07-4}\ (4.1),
(4.2). These formulas have to be generalized for time-dependent processes:

\begin{equation}
R_{abcd}^{(1)}(\mathbf{x},\mathbf{x}^{\prime },t,t^{\prime })=-\beta
\int_{0}^{\infty }dt^{\prime \prime }\int d\mathbf{x}^{\prime \prime }\frac{%
\delta R_{abcd}(\mathbf{x},\mathbf{x}^{\prime },t,t^{\prime })}{\delta b_{e}(%
\mathbf{x}^{\prime \prime },t^{\prime \prime })}|_{\mathbf{u}=0}\,u_{e}(%
\mathbf{x}^{\prime \prime },t^{\prime \prime })  \label{3.1}
\end{equation}

$b_{e}=-\beta \,u_{e}$\ is the momentum part of the conjugated parameters in
the formula for the local equilibrium probability density, see \cite{pi07-4}%
\ (3.6). The functional derivative is calculated in the appendix.

The second-order term of the friction force consists of a parametric and a
functional part:

\begin{equation}
D_{2}=D_{2p}+D_{2f}  \label{3.2}
\end{equation}

When the parametric part of the derivative (\ref{A7}) is inserted into (\ref%
{3.1}), and the result ist introduced into (\ref{2.2}), one obtains the
parametric part of the second-order friction force. We find:%
\begin{equation}
(D_{2p})_{a}(\mathbf{x},t)=\beta ^{2}\nabla _{c}\int d\mathbf{x}^{\prime
}\int d\mathbf{x}^{\prime \prime }\int_{0}^{t}dt^{\prime }\,\,\langle
\lbrack \func{e}^{(1-\mathcal{P}_{0})\mathcal{L}(t-t^{\prime })}(\hat{s}%
_{0})_{ac}(\mathbf{x})](\hat{s}_{0})_{bd}(\mathbf{x}^{\prime })p_{e}(\mathbf{%
x}^{\prime \prime })\rangle _{0}u_{e}(\mathbf{x}^{\prime \prime },t^{\prime
})\nabla _{d}^{\prime }u_{b}(\mathbf{x}^{\prime },t^{\prime })  \label{3.3}
\end{equation}

$\langle \rangle _{0}$ denotes total equilibrium expectation; $\hat{s}_{0}$\
is the flux variable projected with $(1-\mathcal{P}_{0})$; $\mathcal{P}_{0}$%
\ is the total equilibrium counterpart of \ $\mathcal{P}$\ (\ref{2.5}). - If
one inserts the 4$^{th}$ part of the derivative (\ref{A16}) into (\ref{3.1}%
), one obtains the functional part of the friction force:%
\begin{align}
(D_{2f})& _{a}(\mathbf{x},t)=\beta ^{2}\nabla _{c}\int d\mathbf{x}^{\prime
}\int d\mathbf{x}^{\prime \prime }\,\int_{0}^{t}dt^{\prime }\int_{t^{\prime
}}^{t}dt^{\prime \prime }\times   \notag \\
& \times \langle \lbrack \frac{d}{dt^{\prime }}\func{e}^{(1-\mathcal{P}_{0})%
\mathcal{L}(t^{\prime \prime }-t^{\prime })}\mathcal{P}_{0}p_{e}(\mathbf{x}%
^{\prime \prime })\func{e}^{(1-\mathcal{P}_{0})\mathcal{L}(t-t^{\prime
\prime })})(\hat{s}_{0})_{ac}(\mathbf{x})](\hat{s}_{0})_{bd}(\mathbf{x}%
^{\prime })\rangle _{0}\,u_{e}(\mathbf{x}^{\prime \prime },t^{\prime \prime
})\nabla _{d}^{\prime }u_{b}(\mathbf{x}^{\prime },t^{\prime })  \label{3.4}
\end{align}

It is necessary to derive formulas for the appearing correlation functions\
in order to get a definite result. - We begin with transferring the formulas
into Fourier space. In addition, within the correlation formula, we switch
to orthonormal variables $h_{e}$, $r_{ac},$ $\hat{r}_{ac}$\ (see \cite%
{pi07-4}\ (4.7)) .

The hydrodynamic velocity varies over much larger time intervals than the
correlation functions. Therefore, a Markovian approximation of the process
is applied which results in the approximation for the kernel function $f(t)$%
\ in the integrals: $f(t)\rightarrow \delta (t)\int_{0}^{\infty }dt^{\prime
}\,f(t^{\prime })$ . For the parametric part, we obtain:

\begin{equation}
(D_{2p})_{1}(t)=\beta ^{\frac{1}{2}}\rho ^{\frac{3}{2}}\frac{1}{(2\pi )^{6}}%
\int d\mathbf{q}\int d\mathbf{q}^{\prime }(N_{p})_{123}u_{2}(t)u_{3}(t)
\label{3.5}
\end{equation}

\begin{equation}
(N_{p})_{123}=\int_{0}^{\infty }dt^{\prime }\langle \lbrack \func{e}^{(1-%
\mathcal{P}_{0})\mathcal{L}t^{\prime }}(i\mathbf{k}\hat{r})_{1}](i\mathbf{q}%
\hat{r})_{2}^{\ast }h_{3}^{\ast }\rangle _{0}  \label{3.6}
\end{equation}

Number indices are introduced which have been used by many authors. The
number is a combination of an index and a wave number. Certain numbers are
reserved for a specific wave number, as is shown in the following table:%
\begin{equation*}
\begin{array}{c}
\mathbf{k}\text{ : 1, 4, 7,}\cdots \\ 
\mathbf{q}\text{ : 2, 5, 8, }\cdots \\ 
\mathbf{q}^{\prime }\text{ : 3, 6, 9,}\cdots%
\end{array}%
\end{equation*}

Number indices appearing pairwise include a summation over the original
indices; on the other hand, wave number integration will always be shown
explicitly. In (\ref{3.6}), $(i\mathbf{k}\hat{r})_{1}$\ is written for $i%
\mathbf{k}_{d}\hat{r}_{1d}$. \ - $N_{p}$\ is identical with the quantity $N$%
\ in \cite{pi07-4}\ (4.10), which is calculated there. The result (4.27) is
the ''reduced'' form of the kernel function which means that the factor $%
(2\pi )^{3}\delta (\mathbf{k-q-q}^{\prime })$\ is excluded. The full formula
reads%
\begin{equation}
(N_{p})_{123}=(2\pi )^{3}\delta (\mathbf{k-q-q}^{\prime })\frac{1}{2}%
ik_{d}S_{123d}  \label{3.7}
\end{equation}

Details of the matrix $S$\ are not needed and are therefore not repeated
here. - (\ref{3.4}) changes to:

\begin{eqnarray}
(D_{2f})_{1}(t) &=&-\beta ^{\frac{1}{2}}\rho ^{\frac{3}{2}}\frac{1}{(2\pi
)^{6}}\int d\mathbf{q}\int d\mathbf{q}^{\prime
}k_{c}\,q_{d}\int_{0}^{t}dt^{\prime }\int_{t^{\prime }}^{t}dt^{\prime \prime
}\times   \notag \\
&&\times \langle \lbrack \frac{d}{dt^{\prime }}\func{e}^{(1-\mathcal{P}_{0})%
\mathcal{L}(t^{\prime \prime }-t^{\prime })}\mathcal{P}_{0}h_{3}^{\ast }%
\func{e}^{(1-\mathcal{P}_{0})\mathcal{L}(t-t^{\prime \prime })}\hat{r}_{1c}]%
\hat{r}_{2d}^{\ast }\rangle _{0}u_{2}(t^{\prime })u_{3}(t^{\prime \prime })
\label{3.8}
\end{eqnarray}

The projection $\mathcal{P}_{0}$\ in the correlation is performed:

\begin{eqnarray}
(D_{2f})_{1}(t) &=&-\beta ^{\frac{1}{2}}\rho ^{\frac{3}{2}}\frac{1}{(2\pi
)^{6}}\int d\mathbf{q}\int d\mathbf{q}^{\prime }\int_{0}^{t}dt^{\prime
}\int_{t^{\prime }}^{t}dt^{\prime \prime }\times   \notag \\
&&\times \langle \lbrack \func{e}^{(1-\mathcal{P}_{0})\mathcal{L}%
(t-t^{\prime \prime })}(i\mathbf{k}\hat{r})_{1}]h_{2}^{\ast }h_{3}^{\ast }%
\tilde{\rangle}_{0}\langle \lbrack \frac{d}{dt^{\prime }}\func{e}^{(1-%
\mathcal{P}_{0})\mathcal{L}(t^{\prime \prime }-t^{\prime })}h_{2}](i\mathbf{q%
}\hat{r})_{5}^{\ast }\rangle _{0}u_{5}(t^{\prime })u_{3}(t^{\prime \prime })
\label{3.9}
\end{eqnarray}

$\langle \tilde{\rangle}_{0}$ denotes the 3-point-correlation without the
factor $(2\pi )^{3}\delta (\mathbf{k-q-q}^{\prime })$, which we call the
reduced correlation function. - Finally, symbols are introduced for the
correlations appearing in (\ref{3.6}):

\begin{equation}
(D_{2f})_{1}(t)=-\beta ^{\frac{1}{2}}\rho ^{\frac{3}{2}}\frac{1}{(2\pi )^{6}}%
\int d\mathbf{q}\int d\mathbf{q}^{\prime }\int_{0}^{t}dt^{\prime \prime
}\int_{0}^{t^{\prime \prime }}dt^{\prime }(K_{3}\tilde{)}_{123}(t-t^{\prime
\prime })\frac{d(K_{2})_{25}(t^{\prime \prime }-t^{\prime })}{dt^{\prime }}%
u_{3}(t^{\prime \prime })u_{5}(t^{\prime })  \label{3.10}
\end{equation}

\begin{equation}
(K_{3})_{123}(t)=\mathcal{\langle }[\func{e}^{(1-\mathcal{P}_{0})\mathcal{L}%
t}(i\mathbf{k}\hat{r})_{1}]h_{2}^{\ast }h_{3}^{\ast }\rangle _{0}
\label{3.11}
\end{equation}

\begin{equation}
(K_{2})_{25}(t)=\langle \lbrack \func{e}^{(1-\mathcal{P}_{0})\mathcal{L}%
t}h_{2}](i\mathbf{q}\hat{r})_{5}^{\ast }\rangle _{0}  \label{3.12}
\end{equation}

\section{Correlation function $K_{2}$}

\setcounter{equation}{0}

We start the inverstigation of $K_{2}$\ by calculating its initial value: 
\begin{equation}
(K_{2})_{25}(0)=\langle h_{2}(i\mathbf{q}\hat{r})_{5}^{\ast }\rangle
_{0}=\langle \lbrack (1-\mathcal{P})h_{2}](i\mathbf{q}\hat{r})_{5}^{\ast
}\rangle _{0}=0  \label{4.1}
\end{equation}

Here we have used the general properties $\langle f(1-\mathcal{P})g\rangle
_{0}=\langle \lbrack (1-\mathcal{P})f](1-\mathcal{P})g\rangle _{0}$\ and $(1-%
\mathcal{P})h=0$\ . Next we derive a relation between $K_{2}$ and
correlation functions which are defined with the non-projected exponential
operator $\func{e}^{\mathcal{L}t}$. For $K_{2}$, we use the relation:

\begin{equation}
(i\mathbf{k}\hat{r})_{1}=-\dot{h}_{1}-i\,\omega _{14}h_{4}  \label{4.1a}
\end{equation}

We obtain the decomposition:

\begin{eqnarray}
(K_{2})_{25}(t) &=&-\langle \lbrack \func{e}^{(1-\mathcal{P}_{0})\mathcal{L}%
t}h_{2}]\dot{h}_{5}^{\ast }\rangle _{0}+i\omega _{58}\langle \lbrack \func{e}%
^{(1-\mathcal{P}_{0})\mathcal{L}t}h_{2}]h_{8}^{\ast }\rangle _{0}  \notag \\
&&-(K_{21})_{25}(t)+i\omega _{58}(K_{22})_{28}(t)  \label{4.2}
\end{eqnarray}

$\omega _{25}=\langle (\mathbf{q}\hat{r})_{2}h_{5}^{\ast }\rangle _{0}$, and
the symbols in the second row are denotations for the quantities in the
first. \ We use the operator identity \cite{pi07-4} (4.11):

\begin{equation}
\func{e}^{(1-\mathcal{P}_{0})\mathcal{L}t}=\func{e}^{\mathcal{L}%
t}-\int_{0}^{t}dt^{\prime }\func{e}^{\mathcal{L}t^{\prime }}\mathcal{P}_{0}%
\mathcal{L}\func{e}^{(1-\mathcal{P}_{0})\mathcal{L(}t-t^{\prime })}
\label{4.3}
\end{equation}

Application to $K_{21}$\ yields:

\begin{eqnarray}
(K_{21})_{25}(t) &=&-\int_{0}^{t}dt^{\prime }\,\langle \lbrack \func{e}^{%
\mathcal{L}t^{\prime }}\mathcal{P}_{0}\mathcal{L}\func{e}^{(1-\mathcal{P}%
_{0})\mathcal{L(}t-t^{\prime })}h_{2}]\dot{h}_{5}^{\ast }\rangle
_{0}+\langle \lbrack \func{e}^{\mathcal{L}t}h_{2}]\dot{h}_{5}^{\ast }\rangle
_{0}  \notag \\
&=&-\int_{0}^{t}dt^{\prime }(K_{21})_{28}(t-t^{\prime })\frac{d(C_{2}\tilde{)%
}_{85}(t^{\prime })}{dt^{\prime }}-\frac{d(C_{2})_{25}(t)}{dt}  \label{4.4}
\end{eqnarray}

This is an integral equation for $K_{21}$\ in terms of the non-projected
duple correlation\ $(C_{2})_{25}(t)=\langle \lbrack \func{e}^{\mathcal{L}%
t}h_{2}]h_{5}^{\ast }\rangle _{0}$.$\ (C_{2}\tilde{)}$ means $(C_{2})$\
without the factor $(2\pi )^{3}\delta (\mathbf{k-q})$, which we call the
reduced form of the two-point correlation.$\ C_{2}$ obeys the equation \cite%
{pi07-4}\ (4.24):%
\begin{equation}
\frac{d(C_{2})_{25}(t)}{dt}=-\kappa _{28}(C_{2})_{85}(t)\text{ \ , \ \ \ }t>0
\label{4.5}
\end{equation}%
\begin{equation}
\kappa _{28}=i\omega _{25}+\gamma _{25}  \label{4.6}
\end{equation}

$\gamma $ is the dissipation matrix, the space-time integral over the memory
function of the process which is defined with the linear projection operator
employed here. In \cite{slo}\ it is emphasized in connection with formula
(3.18) there, that in general one has to distinguish this from the
corresponding quantities defined with the multilinear projection operator
defined there. For the limited purpose of the present paper, this difference
can be ignored. The condition $t>0$\ in (\ref{4.5}) is essential, since at $%
t=0$\ the time derivative of $C_{2}$\ is discontinuous: While from the
definition of $C_{2}$\ and the conservation relations we find $\frac{d(C_{2}%
\tilde{)}_{25}}{dt}|_{t=0}=-\,i\omega _{25}$, from (\ref{4.5}) one concludes 
$\frac{d(C_{2}\tilde{)}_{25}}{dt}|_{t=0_{+}}=-\,\kappa _{25}$. - (\ref{4.5})
is introduced into (\ref{4.4}), and the integration parameter $t^{\prime }$\
is \ transformed:%
\begin{equation}
(K_{21})_{25}(t)=\kappa _{8,11}\int_{0}^{t}dt^{\prime
}(K_{21})_{28}(t^{\prime })(C_{2}\tilde{)}_{11,5}(t-t^{\prime })+\kappa
_{2,11}(C_{2})_{11,5}(t)\text{\ \ , \ \ \ }t>0  \label{4.7}
\end{equation}

Differentiation with respect to time and insertion of (\ref{4.7}),(\ref{4.5}%
) yields:%
\begin{eqnarray}
\frac{d(K_{21})_{25}(t)}{dt} &=&-\kappa _{14,5}\left\{ \left[
(K_{21})_{2,14}(t)-\kappa _{2,11}(C_{2})_{11,14}(t)\right] +\kappa
_{2,11}(C_{2})_{11,14}(t)\right\} +\kappa _{85}(K_{21})_{28}(t)  \notag \\
&=&0\text{\ \ , \ \ \ }t>0  \label{4.8}
\end{eqnarray}

Thus, for $t>0$, $K_{21}$\ is a constant. The value ist found from (\ref{4.7}%
) by taking the time limit $t=0_{+}$:%
\begin{equation}
(K_{21})_{25}=\kappa _{25}\text{ },\text{ \ \ \ \ }t>0  \label{4.9}
\end{equation}

For calculating the properties of $K_{22}$, we need an operator identity
similar to\ (\ref{4.3}):

\begin{equation}
\func{e}^{(1-\mathcal{P})\mathcal{L}t}=\func{e}^{\mathcal{L}%
t}-\int_{0}^{t}dt^{\prime }\func{e}^{(1-\mathcal{P})\mathcal{L}t^{\prime }}%
\mathcal{PL}\func{e}^{\mathcal{L}(t-t^{\prime })}  \label{4.10}
\end{equation}

Both formulas can be found from more general identities in \cite{zu}. The
calculation ist quite similar to that of $K_{21}$, and the result is:

\begin{equation}
\frac{d(K_{22})_{25}(t)}{dt}=0\text{ },\text{ \ \ \ \ }t>0  \label{4.11}
\end{equation}

\begin{equation}
(K_{22})_{25}=\delta _{25}\text{ },\text{ \ \ \ \ }t>0  \label{4.12}
\end{equation}

Introducing these results into (\ref{4.2}), we obtain:

\begin{equation}
\frac{d(K_{2})_{25}(t)}{dt}=0\text{ },\text{ \ \ \ \ }t>0  \label{4.13}
\end{equation}

\begin{equation}
(K_{2})_{25}=-\gamma _{25}\text{ },\text{ \ \ \ \ }t>0  \label{4.14}
\end{equation}

Combining (\ref{4.14}) and (\ref{4.1}) yields:

\begin{equation}
(K_{2})_{25}(t)=-\gamma _{25}\Theta (t)\text{ },\text{ \ \ \ \ }t\geq 0
\label{4.15}
\end{equation}

$\Theta (t)$\ is the step function. By Differentiation:

\begin{equation}
\frac{d(K_{2})_{25}(t)}{dt}=-\gamma _{25}2\,\delta (t)\text{ },\text{ \ \ \
\ }t\geq 0  \label{4.16}
\end{equation}

It is understood that the dimension of the delta function corresponds to
that of its argument; always the same symbol $\delta $\ is used. The factor
2 is necessary since the delta function has its peak at the boundary of the
definition range of $K_{2}$. - Formula (\ref{4.16}) constitutes the
essential difference to the first version of this paper, where the result
(4.9) stated $\frac{d(K_{2})_{25}(t)}{dt}=0$ $,$ $t\geq 0$. By inserting (%
\ref{4.16}) into (\ref{3.10}) and evaluating the delta function:

\begin{equation}
(D_{2f})_{1}(t)=-\beta ^{\frac{1}{2}}\rho ^{\frac{3}{2}}\frac{\gamma _{25}}{%
(2\pi )^{6}}\int d\mathbf{q}\int d\mathbf{q}^{\prime }\int_{0}^{t}dt^{\prime
}(K_{3})_{153}(t-t^{\prime })u_{2}(t^{\prime })u_{3}(t^{\prime })
\label{4.17}
\end{equation}

Since the hydrodynamic velocity varies on a much longer time scale than the
equilibrium correlation function, it is again possible to apply a Markovian
approximation:

\begin{equation}
(D_{2f})_{1}(t)=\beta ^{\frac{1}{2}}\rho ^{\frac{3}{2}}\frac{1}{(2\pi )^{6}}%
\int d\mathbf{q}\int d\mathbf{q}^{\prime }(N_{f})_{123}u_{2}(t)u_{3}(t)
\label{4.18}
\end{equation}%
\begin{equation}
(N_{f})_{123}=-\gamma _{25}\int_{0}^{\infty }dt^{\prime
}(K_{3})_{153}(t^{\prime })  \label{4.19}
\end{equation}

\section{Calculation of $K_{3}$}

\setcounter{equation}{0}

\bigskip

In contrast to the first version of this paper, we now need to calculate the
correlation function $K_{3}$. We again use the identity\ (\ref{4.3}); with a
calculation which parallels \cite{pi07-4}\ (4.10) to (4.12) we obtain a
formula for $K_{3}$:

\begin{equation}
(K_{3})_{123}(t)=\,-\gamma _{14}\langle \lbrack \func{e}^{\mathcal{L}%
t}h_{4}]h_{2}^{\ast }h_{3}^{\ast }\rangle _{0}+\langle \lbrack \func{e}^{%
\mathcal{L}t}(i\mathbf{k}\hat{r})_{1}]h_{2}^{\ast }h_{3}^{\ast }\rangle _{0}
\label{5.1}
\end{equation}

The first term contains the non-projected single-time triple correlation
function:

\begin{equation}
(C_{3})_{123}(t)=\langle \lbrack \func{e}^{\mathcal{L}t}h_{1}]h_{2}^{\ast
}h_{3}^{\ast }\rangle _{0}  \label{5.2}
\end{equation}

The second term can be transformed by the relation (\ref{4.1a}). We obtain
from (\ref{5.1}):

\begin{equation}
(K_{3}\tilde{)}_{123}(t)=-\kappa _{14}(C_{3}\tilde{)}_{423}(t)-\frac{d(C_{3}%
\tilde{)}_{123}(t)}{dt}  \label{5.3}
\end{equation}

We switched to reduced correlations since this is more suitable for the
following calculation. - As has been stated in \cite{pi07-4}, for $C_{3}$\
expression (4.20) has been derived by multilinear mode coupling theory:

\begin{equation}
(C_{3}\tilde{)}_{123}(t)=(C_{2}\tilde{)}_{14}(t)\,\tilde{J}%
_{423}-\int_{0}^{t}dt^{\prime }(C_{2}\tilde{)}_{14}(t-t^{\prime
})ik_{d}S_{456d}(C_{2}\tilde{)}_{52}(t^{\prime })(C_{2}\tilde{)}%
_{63}(t^{\prime })  \label{5.4}
\end{equation}

$J$ is given in \cite{pi07-4} (4.23). For $K_{3}$, we then find:

\begin{equation}
(K_{3}\tilde{)}_{123}(t)=ik_{d}S_{156d}(C_{2}\tilde{)}_{52}(t)(C_{2}\tilde{)}%
_{63}(t)  \label{5.5}
\end{equation}

The solution of (\ref{4.5}) to the initial condition $(C_{2}\tilde{)}%
_{14}(0)=\delta _{14}$\ is:

\begin{equation}
(C_{2}\tilde{)}_{14}(t)\,=\func{e}^{-\kappa _{14}t}  \label{5.6}
\end{equation}

This is inserted into (\ref{5.5}); by integration we obtain for $N_{f}$\ (%
\ref{4.19}):%
\begin{equation}
(N_{f}\tilde{)}_{123}=-\gamma _{25}(\kappa _{58}\delta _{36}+\delta
_{58}\kappa _{36})^{-1}ik_{d}S_{186d}  \label{5.7}
\end{equation}

For insertion into (\ref{4.18}), we may replace $(N_{f})_{123}$\ by the
symmetric form $\frac{1}{2}((N_{f})_{123}+(N_{f})_{132})$ which we denote by
the same symbol:

\begin{eqnarray}
(N_{f}\tilde{)}_{123} &=&-\frac{1}{2}(\gamma _{25}\delta _{36}+\delta
_{25}\gamma _{36})(\kappa _{58}\delta _{69}+\delta _{58}\kappa
_{69})^{-1}ik_{d}S_{189d}  \notag \\
&=&-\frac{1}{2}ik_{d}S_{123d}+\frac{1}{2}(i\omega _{25}\delta _{36}+\delta
_{25}i\omega _{36})(\kappa _{58}\delta _{69}+\delta _{58}\kappa
_{69})^{-1}ik_{d}S_{189d}  \label{5.8}
\end{eqnarray}

When this is introduced into (\ref{4.18}), it is seen that the second term
vanishes because $\omega _{25}=\breve{\omega}_{5}q_{2}$\ ($\breve{\omega}%
_{5} $ certain constants), and the incompressibility condition\ reads $%
q_{2}u_{2}=0$. Thus, for the relevant part of $N_{f}$, we have:

\begin{equation}
(N_{f})_{123}=-(2\pi )^{3}\delta (\mathbf{k-q-q}^{\prime })\frac{1}{2}%
ik_{d}S_{123d}  \label{5.9}
\end{equation}

When we compare with (\ref{3.7}), we obtain from (\ref{3.2}):

\begin{equation}
D_{2}=0  \label{5.10}
\end{equation}

Thus, with the recalculation of $K_{2}$ (\ref{4.15}), we now obtain the
result that the second-order part of the friction force vanishes. 

\section{Stationary processes}

\setcounter{equation}{0}

\bigskip

We shortly turn to the earlier paper \cite{pi07-4}, where we considered
stationary processes. When the calculation is performed in Fourier space and
orthonormal variables are introduced, inserting (A9) into (4.1) and
importing this into (3.2) of that paper yields the consecutive formula f\"{u}%
r $D_{2f}$:

\begin{equation}
(D_{2f})_{1}=\beta ^{\frac{1}{2}}\rho ^{\frac{3}{2}}\frac{1}{(2\pi )^{6}}%
\int d\mathbf{q}\int d\mathbf{q}^{\prime }k_{c}\,q_{d}\int_{0}^{\infty
}dt^{\prime }\int_{0}^{\infty }dt^{\prime \prime }\langle \lbrack \frac{d}{%
dt^{\prime }}\func{e}^{(1-\mathcal{P})\mathcal{L}t^{\prime }}\mathcal{P}%
h_{3}^{\ast }\func{e}^{(1-\mathcal{P})\mathcal{L}t^{\prime \prime }}\hat{r}%
_{1c}]\hat{r}_{2d}^{\ast }\rangle _{0}u_{2}u_{3}  \label{6.1}
\end{equation}

As in \cite{pi07-4}, the projection operation is performed:

\begin{equation}
(D_{2f})_{1}=\beta ^{\frac{1}{2}}\rho ^{\frac{3}{2}}\frac{k_{c}\,q_{d}}{%
(2\pi )^{6}}\int d\mathbf{q}\int d\mathbf{q}^{\prime }\int_{0}^{\infty
}dt^{\prime }\int_{0}^{\infty }dt^{\prime \prime }\langle \lbrack \func{e}%
^{(1-\mathcal{P})\mathcal{L}t^{\prime \prime }}\hat{r}_{1c}]h_{5}^{\ast
}h_{3}^{\ast }\rangle _{0}\frac{d}{dt^{\prime }}\langle \lbrack \func{e}^{(1-%
\mathcal{P})\mathcal{L}t^{\prime }}h_{5}]\hat{r}_{2d}^{\ast }\rangle
_{0}u_{2}u_{3}  \label{6.2}
\end{equation}

It has been taken into account that we have:

\begin{equation}
\langle \lbrack \func{e}^{(1-\mathcal{P})\mathcal{L}t^{\prime \prime }}\hat{r%
}_{1c}]h_{3}^{\ast }\rangle _{0}=\langle \lbrack \func{e}^{\mathcal{L}(1-%
\mathcal{P})t^{\prime \prime }}\hat{r}_{1c}](1-\mathcal{P})h_{3}^{\ast
}\rangle _{0}=0  \label{6.3}
\end{equation}

Definitions (\ref{3.11}), (\ref{3.12}) of the present paper are introduced:

\begin{eqnarray}
(D_{2f})_{1} &=&-\beta ^{\frac{1}{2}}\rho ^{\frac{3}{2}}\frac{1}{(2\pi )^{6}}%
\int d\mathbf{q}\int d\mathbf{q}^{\prime }u_{2}u_{3}\int_{0}^{\infty
}dt^{\prime \prime }(K_{3})_{153}(t^{\prime \prime })\int_{0}^{\infty
}dt^{\prime }\frac{d(K_{2}\tilde{)}_{52}(t^{\prime })}{dt^{\prime \prime }} 
\notag \\
&=&-\beta ^{\frac{1}{2}}\rho ^{\frac{3}{2}}\frac{1}{(2\pi )^{6}}\int d%
\mathbf{q}\int d\mathbf{q}^{\prime }u_{2}u_{3}\int_{0}^{\infty }dt^{\prime
\prime }(K_{3})_{153}(t^{\prime \prime })\left[ \lim_{t^{\prime }\rightarrow
\infty }(K_{2}\tilde{)}_{52}(t^{\prime })-(K_{2}\tilde{)}_{52}(0)\right] 
\label{6.4}
\end{eqnarray}

It has been assumed in \cite{pi07-4}\ that for large $t$\ the factor
variables of $K_{2}$\ become statistically independent; this led to $%
\lim_{t^{\prime }\rightarrow \infty }(K_{2})_{52}(t^{\prime })=0$. But as we
see from (\ref{4.14}), for $K_{2}$\ this assumption is not true. When the
correct formula is incorporated, the calculation leads to the final result (%
\ref{5.10}).

\section{Summary}

\bigskip 

As is stated in the introduction, $D_{2}=0$ ensures that the Navier-Stokes
equation is obtained correctly as second-order velocity approximation of the
mean momentum equation derived by projection operator technique. 

\appendix

\section{Appendix: Calculation of the functional derivative}

\setcounter{equation}{0}

The kernel function (\ref{2.3}) depends on $b_{e}$\ 4-fold, namely, in the
formula for the local equilibrium density $\rho _{L}$, in $\mathcal{P}$\
contained in $\hat{s}_{ac}$\ and $\hat{s}_{bd}$, and in the operator $%
\mathcal{G}$. For abbreviation, the formula for the derivative is written:

\begin{equation}
\frac{\delta R_{abcd}(%
\vekt{x}%
,%
\vekt{x}%
^{\prime },t,t^{\prime })}{\delta b_{e}(%
\vekt{x}%
^{\prime \prime },t^{\prime \prime })}=\sum_{i=1}^{4}\left[ \frac{\delta R}{%
\delta b}\right] ^{(i)}  \label{A1}
\end{equation}

In the first three of these, the time dependence of $R$\ is parametric via
the time dependence of $\rho _{L}$, at the time instant $t$\ or $t^{\prime }$%
. In these cases, the time-dependent derivative is equal to the
corresponding stationary derivative times $\delta (t^{\prime \prime }-t)$\
or $\delta (t^{\prime \prime }-t^{\prime })$. The stationary derivatives are
given in \cite{pi07-4} \ (A3) to (A5). We still have to take these formulas
at $\mathbf{u}=0$; then, $\rho _{L}$\ switches to $\rho _{0}$, the
probability density of total equilibrium (for a given density and
temperature), and therefore $\langle \rangle _{L,t}$\ to $\langle \rangle
_{0}$\ , the equilibrium expectation; $\mathcal{P}$\ reduces to $\mathcal{P}%
_{0}$\ , the corresponding total equilibrium projection operator; $\hat{s}%
_{ac}$\ to $(\hat{s}_{0})_{ac}$, the flux density projected by $\mathcal{P}%
_{0}$; $\delta p_{e}$\ to $p_{e}$, and $\mathcal{G}(t^{\prime },t)$\ to $%
\func{e}^{(1-\mathcal{P}_{0})\mathcal{L(}t-t^{\prime })}$. We obtain:

\begin{equation}
\left[ \frac{\delta R_{abcd}}{\delta b_{e}}\right] _{\mathbf{u}%
=0}^{(1)}=-\delta (t^{\prime \prime }-t^{\prime })\beta \,\langle \lbrack 
\func{e}^{\mathcal{L}(1-\mathcal{P}_{0})(t-t^{\prime })}(\hat{s}_{0})_{ac}(%
\mathbf{x})]p_{e}(\mathbf{x}^{\prime \prime })(\hat{s}_{0})_{bd}(\mathbf{x}%
^{\prime })\rangle _{0}  \label{A2}
\end{equation}

\begin{equation}
\left[ \frac{\delta R_{abcd}}{\delta b_{e}}\right] _{\mathbf{u}%
=0}^{(2)}=\delta (t^{\prime \prime }-t^{\prime })\beta \,\langle \lbrack 
\func{e}^{\mathcal{L}(1-\mathcal{P}_{0})(t-t^{\prime })}(\hat{s}_{0})_{ac}(%
\mathbf{x})]\mathcal{P}_{0}p_{e}(\mathbf{x}^{\prime \prime })(\hat{s}%
_{0})_{bd}(\mathbf{x}^{\prime })\rangle _{0}  \label{A3}
\end{equation}

\begin{equation}
\left[ \frac{\delta R_{abcd}}{\delta b_{e}}\right] _{\mathbf{u}%
=0}^{(3)}=\delta (t^{\prime \prime }-t)\beta \,\langle \lbrack \func{e}^{%
\mathcal{L}(1-\mathcal{P}_{0})(t-t^{\prime })}\mathcal{P}_{0}p_{e}(\mathbf{x}%
^{\prime \prime })(\hat{s}_{0})_{ac}(\mathbf{x})](\hat{s}_{0})_{bd}(\mathbf{x%
}^{\prime })\rangle _{0}  \label{A4}
\end{equation}

We have:%
\begin{eqnarray}
\langle \lbrack \func{e}^{\mathcal{L}(1-\mathcal{P}_{0})t}\mathcal{P}%
_{0}A](1-\mathcal{P}_{0})B\rangle _{0} &=&\langle \lbrack (1-\mathcal{P}_{0})%
\func{e}^{\mathcal{L}(1-\mathcal{P}_{0})t}\mathcal{P}_{0}A](1-\mathcal{P}%
_{0})B\rangle _{0}  \notag \\
&=&\langle \lbrack \func{e}^{(1-\mathcal{P}_{0})\mathcal{L}t}(1-\mathcal{P}%
_{0})\mathcal{P}_{0}A](1-\mathcal{P}_{0})B\rangle _{0}  \notag \\
&=&0  \label{A5}
\end{eqnarray}

From this we find:

\begin{equation}
\left[ \frac{\delta R_{abcd}}{\delta b_{e}}\right] _{\mathbf{u}=0}^{(3)}=0
\label{A6}
\end{equation}

The remaining two parts yield the parametric part of the functional
derivative:%
\begin{eqnarray}
\left[ \frac{\delta R_{abcd}}{\delta b_{e}}\right] _{p} &=&\left[ \frac{%
\delta R_{abcd}}{\delta b_{e}}\right] _{\mathbf{u}=0}^{(1)}+\left[ \frac{%
\delta R_{abcd}}{\delta b_{e}}\right] _{\mathbf{u}=0}^{(2)}  \notag \\
&=&-\delta (t^{\prime \prime }-t^{\prime })\beta \,\langle \lbrack \func{e}^{%
\mathcal{L}(1-\mathcal{P}_{0})(t-t^{\prime })}(\hat{s}_{0})_{ac}(\mathbf{x}%
)](1-\mathcal{P}_{0}\mathcal{)}p_{e}(\mathbf{x}^{\prime \prime })(\hat{s}%
_{0})_{bd}(\mathbf{x}^{\prime })\rangle _{0}  \notag \\
&=&-\delta (t^{\prime \prime }-t^{\prime })\beta \,\langle \lbrack \func{e}%
^{(1-\mathcal{P}_{0})\mathcal{L}(t-t^{\prime })}(\hat{s}_{0})_{ac}(\mathbf{x}%
)](\hat{s}_{0})_{bd}(\mathbf{x}^{\prime })p_{e}(\mathbf{x}^{\prime \prime
})\rangle _{0}  \label{A7}
\end{eqnarray}

Note that in the 2$^{nd}$ step, in the exponent the sequence of operators is
reversed; compare (\ref{A5}) for a similar operation.

The 4$^{th}$\ part of the derivative is defined:

\begin{equation}
\left[ \frac{\delta R_{abcd}}{\delta b_{e}(\mathbf{x}^{\prime \prime
},t^{\prime \prime })}\right] ^{(4)}=\beta \langle \lbrack \frac{\delta 
\mathcal{G}(t^{\prime },t)}{\delta b_{e}(\mathbf{x}^{\prime \prime
},t^{\prime \prime })}(1-\mathcal{P}(t))s_{ac}(\mathbf{x})](1-\mathcal{P}%
(t^{\prime }))s_{bd}(\mathbf{x}^{\prime })\rangle _{L,t^{\prime }}
\label{A8}
\end{equation}

By temporarily approximating the Integral in (\ref{2.6}) by a sum, one finds
for the derivative of $\mathcal{G}(t^{\prime },t)$:

\begin{equation}
\frac{\delta \mathcal{G}(t^{\prime },t)}{\delta b_{e}(\mathbf{x}^{\prime
\prime },t^{\prime \prime })}=\int_{t^{\prime }}^{t}d\tau \mathcal{G}%
(t^{\prime },\tau )\frac{\delta (\mathcal{L}(1-\mathcal{P}(\tau )\mathcal{)})%
}{\delta b_{e}(\mathbf{x}^{\prime \prime },t^{\prime \prime })}\mathcal{G}%
(\tau ,t)  \label{A9}
\end{equation}

Moreover, for the middle factor of the integrand which depends
parametrically on $\tau $, we have:

\begin{equation}
\frac{\delta (\mathcal{L}(1-\mathcal{P}(\tau )\mathcal{)})}{\delta b_{e}(%
\mathbf{x}^{\prime \prime },t^{\prime \prime })}=\delta (t^{\prime \prime
}-\tau )\mathcal{LP}(t^{\prime \prime }\mathcal{)}\delta p_{e}(\mathbf{x}%
^{\prime \prime },t^{\prime \prime })(1-\mathcal{P}(t^{\prime \prime }%
\mathcal{)})  \label{A10}
\end{equation}

This is introduced into (\ref{A9}). The following structure arises:

\begin{equation}
\int_{t^{\prime }}^{t}d\tau \delta (t^{\prime \prime }-\tau )F(t^{\prime
\prime },\tau )=\Theta (t^{\prime \prime }-t^{\prime })\Theta (t-t^{\prime
\prime })F(t^{\prime \prime },t^{\prime \prime })  \label{A11}
\end{equation}

with $\Theta (t)$\ being the step function:%
\begin{equation}
\Theta (t)=\left\{ 
\begin{array}{c}
0,\quad t\leq 0 \\ 
1,\quad t>0%
\end{array}%
\right.   \label{A12}
\end{equation}

We obtain:

\begin{equation}
\frac{\delta \mathcal{G}(t^{\prime },t)}{\delta b_{e}(\mathbf{x}^{\prime
\prime },t^{\prime \prime })}=\Theta (t^{\prime \prime }-t^{\prime })\Theta
(t-t^{\prime \prime })\mathcal{G}(t^{\prime },t^{\prime \prime })\mathcal{LP}%
(t^{\prime \prime }\mathcal{)}\delta p_{e}(\mathbf{x}^{\prime \prime
},t^{\prime \prime })(1-\mathcal{P}(t^{\prime \prime }\mathcal{)})\mathcal{G}%
(t^{\prime \prime },t)  \label{A13}
\end{equation}

This is introduced into (\ref{A8}):

\begin{align}
\left[ \frac{\delta R_{abcd}}{\delta b_{e}(\mathbf{x}^{\prime \prime
},t^{\prime \prime })}\right] ^{(4)}& =\beta \Theta (t^{\prime \prime
}-t^{\prime })\Theta (t-t^{\prime \prime })\times  \notag \\
& \times \langle \lbrack \mathcal{G}(t^{\prime },t^{\prime \prime })\mathcal{%
LP}(t^{\prime \prime }\mathcal{)}\delta p_{e}(\mathbf{x}^{\prime \prime
},t^{\prime \prime })(1-\mathcal{P}(t^{\prime \prime }\mathcal{)})\mathcal{G}%
(t^{\prime \prime },t)\hat{s}_{ac}(\mathbf{x},t)]\hat{s}_{bd}(\mathbf{x}%
^{\prime },t^{\prime })\rangle _{L,t^{\prime }}  \label{A14}
\end{align}

For $\mathbf{u}=0$, this reads:

\begin{align}
\left[ \frac{\delta R_{abcd}}{\delta b_{e}(\mathbf{x}^{\prime \prime
},t^{\prime \prime })}\right] _{\mathbf{u}=0}^{(4)}& =\beta \Theta
(t^{\prime \prime }-t^{\prime })\Theta (t-t^{\prime \prime })\times  \notag
\\
& \times \langle \lbrack \func{e}^{\mathcal{L}(1-\mathcal{P}_{0})(t^{\prime
\prime }-t^{\prime })}\mathcal{LP}_{0}p_{e}(\mathbf{x}^{\prime \prime })(1-%
\mathcal{P}_{0})\func{e}^{\mathcal{L}(1-\mathcal{P}_{0})(t-t^{\prime \prime
})})(\hat{s}_{0})_{ac}(\mathbf{x})](\hat{s}_{0})_{bd}(\mathbf{x}^{\prime
})\rangle _{0}  \label{A15}
\end{align}

After some manipulations, the result is:

\begin{align}
\left[ \frac{\delta R_{abcd}}{\delta b_{e}(\mathbf{x}^{\prime \prime
},t^{\prime \prime })}\right] _{\mathbf{u}=0}^{(4)}& =\beta \Theta
(t^{\prime \prime }-t^{\prime })\Theta (t-t^{\prime \prime })\times  \notag
\\
& \times \langle \lbrack \func{e}^{(1-\mathcal{P}_{0})\mathcal{L}(t^{\prime
\prime }-t^{\prime })}(1-\mathcal{P}_{0})\mathcal{LP}_{0}p_{e}(\mathbf{x}%
^{\prime \prime })\func{e}^{(1-\mathcal{P}_{0})\mathcal{L}(t-t^{\prime
\prime })})(\hat{s}_{0})_{ac}(\mathbf{x})](\hat{s}_{0})_{bd}(\mathbf{x}%
^{\prime })\rangle _{0}  \notag \\
& =-\beta \Theta (t^{\prime \prime }-t^{\prime })\Theta (t-t^{\prime \prime
})\times  \notag \\
& \times \langle \lbrack \frac{d}{dt^{\prime }}\func{e}^{(1-\mathcal{P}_{0})%
\mathcal{L}(t^{\prime \prime }-t^{\prime })}\mathcal{P}_{0}p_{e}(\mathbf{x}%
^{\prime \prime })\func{e}^{(1-\mathcal{P}_{0})\mathcal{L}(t-t^{\prime
\prime })})(\hat{s}_{0})_{ac}(\mathbf{x})](\hat{s}_{0})_{bd}(\mathbf{x}%
^{\prime })\rangle _{0}  \label{A16}
\end{align}

\end{document}